\documentclass{aa}  

\def\Msun{M$_\odot$}
\def\msun{M$_\odot$}

\def\Mv{M$_{\rm v}$} 
\def\Teff{T$_{\rm eff}$} 
 
\def\0BMV{(B--V)$_{\rm 0}$} 
\def\BMV{B--V} 
\def\Z{Z} 
\def\Y{Y}
\def\V{V}

\def\simgt{\lower.5ex\hbox{$\; \buildrel > \over \sim \;$}} 
\def\simlt{\lower.5ex\hbox{$\; \buildrel < \over \sim \;$}}
\def\ocen{$\omega$~Cen}

\usepackage{graphicx}
\usepackage{natbib}
\usepackage{txfonts}
\begin{document}
   \title{NGC 6441: another indication for a very high helium content in
Globular Cluster stars.}


   \author{V. Caloi
          \inst{1}
          \and
          F. D'Antona\inst{2}
          }

   \offprints{V. Caloi}

   \institute{INAF -Istituto di Astrofisica Spaziale e Fisica 
             Cosmica-Roma,  Via Fosso del Cavaliere, 
             00133 Roma, Italy\\
              \email{vittoria.caloi@iasf-roma.inaf.it}
         \and
             {INAF - Osservatorio Astronomico  di Roma, INAF, via 
             Frascati 33, 00040 Monte Porzio, Italy\\
             \email{dantona@oa-roma.inaf.it}
             }}


 
  \abstract
  {The metal-rich bulge globular cluster NGC 6441 shows a well
developed blue horizontal branch (Rich et al.), together with a
strong slope upward from the red clump to the blue of the RR Lyrae
region. Both features, the former corresponding to the well-known second
parameter problem, are not explained by conventional evolutionary models.}
   { Helium self-enrichment is proposed as a possible solution to both
questions, a mechanism already invoked for the interpretation of the
peculiarities in NGC 2808 and M13. }
   {We make use of horizontal branch simulations, covering a wide
range in main sequence helium abundance, to investigate whether the main
features of NGC 6441 horizontal branch population, including the RR
Lyrae variables period, can be reproduced. }
 {To describe the horizontal branch of NGC 6441, the helium content Y
  in the red clump must reach at least 0.35; values up to Y$\sim$0.37
are necessary to populate the RR Lyr region, reproducing also the
observed mean long periods; depending on the dispersion in mass loss
 assumed in the simulations, values up to Y$\sim$0.38--0.40 are necessary
  to populate the blue HB. The total self--enriched population amounts
  to $\sim$60\% of the whole stellar content.}
   { Self-enrichment and multiple star formation episodes in the
early evolution of globular clusters appear more and more able to
account for many of the chemical and population peculiarities observed
in these systems. The very large helium abundances (Y$\simgt$0.35)
required for $\sim$14\% of the horizontal branch population  pose some problem
  on the enrichment mechanisms.}


\authorrunning{Caloi and D'Antona}

\titlerunning{High helium in NGC 6441}
  
   \maketitle
%

\section{Introduction}

In the last few years, some of the (by now) well known abundance
anomalies in globular cluster (GC) star composition have been observed
in unevolved cluster members (f.e., the anticorrelation between Na and O
abundances; Gratton et al. 2001, Ramirez \& Cohen 2003, Cohen \&
Melendez 2005, Carretta et al. 2006). These observations have focused
attention on the early stages of GC formation and evolution, since it
appears necessary that the abundance deviations from that of the
pristine material have taken place before the formation of the stars in
which they are observed.

A model which can explain all the chemical anomalies is not yet
available, but some aspects of the problem begin to be understood,
while others are still obscure. A clear step forward was made when
it was recognized the link between the chemical anomalies and the
horizontal branch (HB) morphology of the clusters: if the chemical
anomalies, and in particular the Na--O anticorrelation, is interpreted
in terms of a second generation of stars born from the ejecta of the
massive asymptotic giant branch (AGB) stars of the first stellar
generation, there must also be a helium discontinuity between the two
samples of stars. In fact the AGB ejecta from which the second
generation is born are all helium enriched\footnote{The masses of the
AGBs involved must be $\simgt$4\msun, otherwhise the second generation
would show evolutionary signatures of a prolonged phase of third dredge
up, namely the sum of CNO abundances would not be costant, and there
would be s-- process elements enhancement. The helium content of the
ejecta of a 4 \Msun\ AGB star is Y$\sim$0.27 for Z=0.001.}
\citep{dantona2002} and consequently the second star generation evolving
along the HB have bluer locations with respect to the first generation
HB stars. This model was able to produce a detailed explanation of the
anomalous morphology of the HB of the cluster NGC~2808
\citep{dantonacaloi2004}; besides, observations have confirmed the
existence of a Na--O anticorrelation in the cluster
\citep{carretta28082006}. 

An intriguing problem concerning at the moment only NGC~2808 and the
``cluster" \ocen\, is the presence of a well defined second main
sequence (MS), bluer than the main one. In terms of standard MS models,
this feature is a helium rich sequence in which the helium mass fraction
must be Y$\sim$0.4 (see for \ocen\ \citealt{norris, piotto2005} and for
NGC~2808 \citealt{dantona2005}).  AGB models do not predict such a large
helium abundance in the ejecta, the maximum computed value being
Y$\simeq$0.36 in \cite{lattanzio2004}, obtained from the evolution of a
6\msun\ with Z=0.004. Ventura et al. (2002) quote values up to
Y$\simeq$0.31 and Ventura \& D'Antona (2005b) estimate Y=0.32 for
their most massive models (6.5 \msun) for metallicity Z=0.001. Although
this value can be considered conservative, as these models do not
include any kind of overshooting, a preliminary exploration of the
second dredge up does not increase Y beyond $\sim 0.35$ (Ventura 2006,
private communication). The episodes of third dredge up may help to
increase the helium content in some models \citep{venturadantona2005a},
but too many episodes in the end do not preserve the constancy of
C+N+O. Inclusion of the results from these computation in a code for
computing the chemical evolution of helium confirms the conclusion that
present AGB models cannot explain the extreme helium enhancements
\citep{karakas2006}.
The role of super--AGBs, the stars which ignite Carbon in semidegenerate 
conditions and may evolve through a AGB phase, ending up their life as
O--Ne--Mg white dwarfs, has not yet been explored for the metallicities of GCs. 
\cite{siess} finds Y $\sim$ 0.38 for these stars after the second
dredge up, but this result may be affected by the initial value of
the helium content, as his models refer to solar metallicity.

From \cite{dantona2005}, the extreme helium enhancement, derived both
from the blue MS and from the extreme HB blue tails EBT2 and EBT3 (as
defined in \citealt{bedin2000}), concerns only $\sim$15\% of the stars in
NGC~2808, and a similar fraction of stars is involved in the bluer MS in
\ocen. 

We remind that the extreme value of the helium content, obtained
from the MS fit, is rather uncertain. 
On the basis of our present understanding, extreme helium
enhancements (Y $>$ 0.35) may derive either from the high mass tail of
AGBs (for which at present we do not yet have models, see above) or from
the wind ejecta of the massive stars of the first stellar generation
\citep{piotto2005, prantzos2006, smith2006}.  This latter suggestion
would complicate even more the picture of star formation in the
``simple" stellar population such as GC stars were considered until a
few years ago.
It is then important to ascertain whether there are any other GC in
which there is a clear indication for an extreme helium rich (\Y$>$0.35)
component. In the absence of new information coming from other main
sequences, we considered modelling the peculiar HB of the cluster
NGC~6441, whose morphology and RR Lyr population (Rich et al. 1997,
Layden et al 1999, Pritzl et al. 2000) are both a clear
indication of the presence of helium enrichment, as already suggested by
\cite{sweigart1998, dantona2002, dantonacaloi2004}. The results obtained from
this analysis can be extended to the cluster NGC 6388, very similar to 
NGC 6441 in metallicity and peculiar HB morphology and RR Lyr periods.
 
We shall proceed as follows: \S 2 presents the observational data for NGC
6441 and the related problems; the theoretical background is given in
\S 3, while \S 4 describes the comparison with the observations with a
possible solution to the questions in \S 2; some aspects of the given
solution are examined in \S 5, and \S6 gives the final discussion.

   \begin{figure}
   \centering
   \includegraphics[width=8cm]{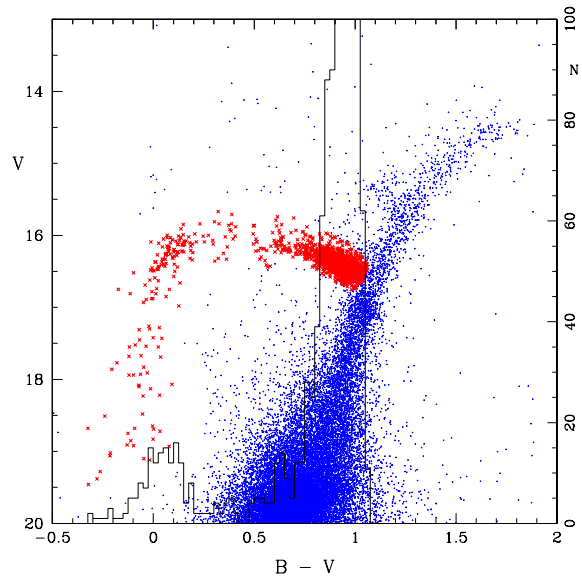}
      \caption{CM diagram of NGC~6441 from the data by Piotto et
    al. 2002. The crosses (in red) indicate our selection of HB
    members; also shown is the histogram of their distribution with respect 
    to \BMV}
         \label{fig1}
   \end{figure}
\begin{figure}
   \centering
   \includegraphics[width=8cm]{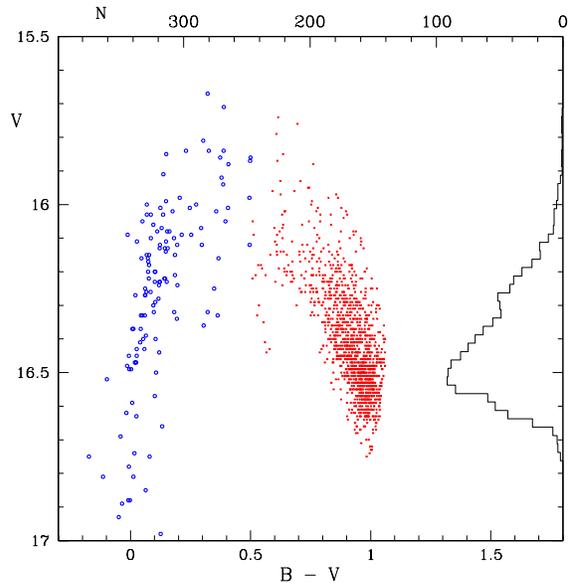}
      \caption{An enlargement of the HB red clump region: crosses (red)
      indicate stars with
     \BMV\ $>$ 0.5~mag, open circles (blue) stars with \BMV\ $<$ 0.5~mag.
     The histogram of red HB stars with respect to V mag is shown.}
         \label{fig2}
   \end{figure}

\section{The horizontal branch in NGC 6441}

NGC 6441 belongs to the metal rich group of GCs (Z $\simgt$ 0.004). All
these clusters have a HB confined to the red side of the RR Lyr region,
with the exception of NGC 6441 itself and of its ``twin cluster'' NGC
6388 (Rich et al. 1997).

The CM diagram of NGC 6441 is shown in Fig. \ref{fig1} (data from
\citealt{piotto2002}, dereddened by E(B--V)=0.44). 
The red clump has a strong slope upward, which
continues in the RR Lyrae and blue HB regions, with a difference of at
least 0.6 mag between the top of the blue HB and the bottom of the red
clump. The features to be explained are: a) the presence of HB
stars hotter than the red clump ones; b) the strong tilt in luminosity,
much larger than expected on the basis of evolutionary models (f.e.,
Raimondo et al. 2002); c) the very long average period of the RR Lyr
variables, unexpected on the basis of the metal content (Z $\ge$ 0.006;
Layden et al. 1999, Pritzl et al. 2001, 2003); d) the presence of an
extended blue tail down to a magnitude V $\sim$ 19.5, corresponding to
\Mv\ $\sim$ 3.5 mag.

In Fig. \ref{fig1} our choice of the HB members is indicated
(crosses). The numbers turn out to be: a total of 1452 members, out of
which 1278 form a red clump (\0BMV\ $>$ 0.5). For our simulations we
assumed therefore a total of 1452 stars, to be distributed according to
observations. Also in Fig. \ref{fig1} the histogram of the number of HB
stars with respect to \BMV\, is shown; in Fig. \ref{fig2} the histogram
with respect to \V\, mag is shown, considering only stars with \0BMV\,
$>$ 0.5 (crosses).

The number of RR Lyr is 45 in the sample by \cite{piotto2002} (Piotto,
private communication). We assume a mean fundamentalized period (see below)
of 0.67~d, based on 63 RRab and RRc by \cite{pritzl2003}.

\section{The evolutionary and synthetic HB models}

The computation of HB models and of the synthetic HBs follows the
prescription in D'Antona \& Caloi (2004). We will model the HB by assuming
cluster members to be a mixture of stars with different helium
abundances, as we did for NGC~2808 and M13.  Previous simulations for
the HB of NGC~6441 have been performed by \cite{sweigart1998}, who also
considered models with high Y, but not a mixture of helium contents. 

The scheme of our modelling will be the following.  We consider a first
group of stars with a fixed helium abundance \Y, close to the
cosmological value \Y = 0.25, which represents the first stellar
generation. A second (or more) generation(s) are added, with a helium
content variable from \Y$\simgt 0.27$\ to \Y$\simlt$0.40.  The number
vs. helium distribution will be chosen in order to comply with the
relevant HB features mentioned in Sect. 2

\subsection{HB structures}

We computed evolutionary sequences from the main sequence to the red
giant tip for a heavy element content of Z = 0.006 and Y = 0.24, 0.26,
0.30 and 0.40. Having an estimate of the helium core masses at the
flash, we obtained HB evolutionary sequences with the appropriate core
masses and atmospheric helium abundances (after the first
dredge--up). The theoretical tracks were transformed into Johnson magnitudes
B and V by means of \cite{castelli}.
The clump colours correspond nicely to the colours of the transformed 
tracks. Nevertheless, we do not emphasize this correspondence as a
proof of a good choice of metallicity, as we could have shifted the
observed colors within the uncertainties of the reddening ($\pm 0.03$
according to \cite{layden1999}). Such an uncertainty has no influence 
upon our comparisons, based quantitatively on luminosity differences
and periods and only qualitatively on the color extension of the HB, 
which greatly exceeds the reddening error. 
On this basis, HB simulations were computed in \V, \BMV,
assuming a gaussian error on magnitudes and colours. A standard
deviation of 0.05 mag has been chosen on the basis of the HB data 
from Piotto et al. (2002).

\subsection{The main parameters in the simulations} 

We assumed as the basic parameter describing the HB distribution the
ratio R of the number of blue (\0BMV\,$<$ 0.5) HB members to the total
number of HB stars. This ratio (174/1452 with the numbers above) is R =
0.12 $\pm$ 0.01. The other constraints to be satisfied by the
simulations are the agreement with the histogram in Fig. \ref{fig2}, the
extension in magnitude of the blue tail and the mean RR Lyr period.

With regard to the RR Lyr variables, for every simulation we evaluated
the mean period of the artificial RR Lyrae variables by means of
\cite{va1973} formulation; the limits of the RR Lyr region were taken
from log\Teff\ = 3.79 to 3.875. This is a standard interval for the more
metal poor GCs, and may not correspond exactly to the present conditions
of high \Z\, and high luminosity.  The properties of the RR Lyraes in
NGC 6441 appear extreme when compared with those of the whole family of
GCs (see, f.e., Fig.10 in Pritzl et al. 2001), but still within the
limits of current interpretation.  So we are confident that the order of
magnitude of the estimated period is adequate.  The observational mean
period is built by averaging the observed periods, taking care to
``fundamentalize'' the periods of the RRc variables by adding 0.128 to
the logarithm of their periods \citep{va1973}.

\begin{figure}
   \centering
   \includegraphics[width=8cm]{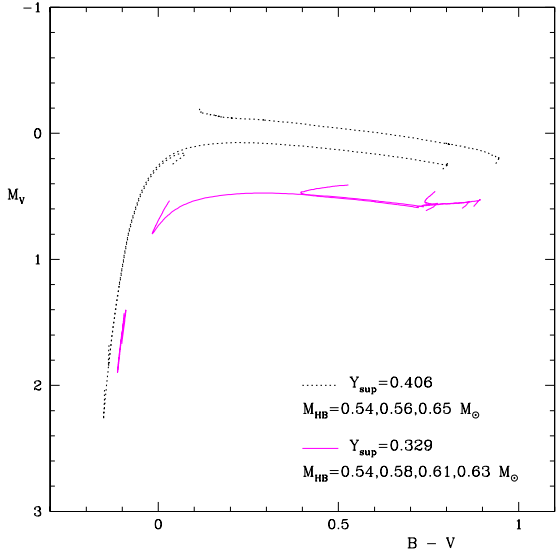}
      \caption{Evolutionary tracks for a few helium rich HB models:
    the long loops that they perform cover large portions
    of the allowed HB region. The tracks end at Y$_{\rm c}$=0.10.}
      \label{fig3}
   \end{figure}
\begin{figure}
   \centering
   \includegraphics[width=8cm]{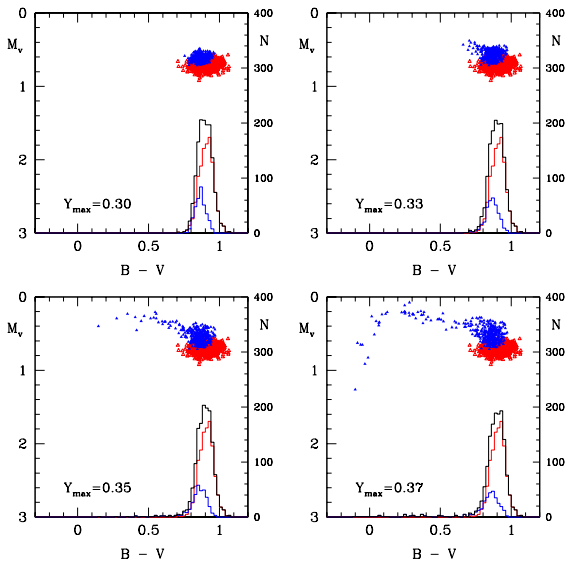}
      \caption{Simulations show
    the effect on the evolutionary tracks of an increasing 
    helium content: open triangles (red) indicate stars with fixed
    Y=0.25 (red histogram), filled triangles (blue) stars with 
    increasing maximum helium content (Y$_{\rm max}$) (blue histogram), 
    as labeled in the four panels. 
    The black histogram refers to the whole HB.  \Z =0.006 in all models.}
         \label{fig4}
   \end{figure}

\section{The simulations and the comparison with the data}

Two features, one observational and the other theoretical, affect
strongly the comparison between observations and simulations:
\begin{enumerate}
\item the fact
that the RR Lyrae variables and the blue component in NGC 6441 are only
a tail of the red clump distribution, covering slighly more than 10\% of
the total HB population;
\item the behaviour in the HR diagram of HB
models rich in helium.
\end{enumerate}
The evolutionary path of the helium rich models may be
confined in the red clump region, or may perform long loops, reaching
very blue colours, for a difference in total mass of the order of $\simlt$0.02
\Msun: a well known fact, see, f.e., \cite{sweigartgross1976}.  
This is shown also in Fig. \ref{fig3}. 
Both these features conspire to vary noticeably the RR Lyr number
and mean period, even by only varying the random extraction.
Consequently, we will try to fit first of all the observed fraction
of stars at \BMV$<$0.5 mag, and will
impose the extension of the blue tail down to \Mv\ $>$ 3 mag. In any case, we
will not accept for the RR Lyrs a mean period smaller than 0.66~d.

\subsection{Age, helium distribution and mass loss}

We assumed a reference age of 11 Gyr. Coeval populations with the same
metallicity but different helium content will have turnoff masses
inversely proportional to Y. Therefore, the progeny of the helium rich,
lower mass evolving red giants will be located, on the average, at
larger temperatures (due to the smaller evolving masses) and larger
luminosity (due to the higher efficiency of the hydrogen shell burning)
than the structures with lower helium.  To illustrate this basic fact,
we show in Fig. \ref{fig4} the behaviour of two groups of stars, in one of
which the maximum helium content is progressively increased. As
expected, the evolutionary paths, confined within the red clump for
lower Y, as it happens for the cosmological helium population, reach
farther and farther into the bluer HB regions when increasing the
maximum Y.

For an age of 11 Gyr, we chose as mass loss during the red giant
evolution $\Delta$M=0.165 \Msun\, with $\sigma$=0.02. With this amount
it was easy to obtain the desired behaviour of the clump luminosity,
when varying the helium content in the helium enriched population(s). A
larger mass loss would cause an excessive loss of clump members
(increasing the fraction of stars at \BMV $<$0.5 mag 
beyond the allowed values), a lower one would have the
opposite effect. Of course the value of 0.165 \Msun\, is related to the
chosen age: for 13 Gyr, a mass loss of 0.13 \Msun\, would have given
similar results.

The faintest blue HB stars in Fig. \ref{fig1} -- about 3.5 mag less
luminous than the RR Lyrs -- correspond to a HB mass $\le$ 0.5
\Msun. For Y = 0.40 and an age of 11 Gyr the evolving red giant has a
mass of $\sim$ 0.72 \Msun, which would require a mass loss $\ge$ 0.22
\Msun; larger mass losses would be necessary for lower Y, up to 0.45
\Msun\ for Y=0.25. We have therefore considered models with Y up to
0.40, but even at this large helium content the very faint magnitudes
are statistically difficult to obtain. So we took into account the
possibility of an increase in the mass loss rate with the decrease of
the red giant mass, according to Reimers (1977) mass loss relation (see
also Lee et al. 1994).  But this is not an automatic solution to the
problem; the progression in mass loss used in Lee et al. (1994) would
not allow any HB progeny to red giants of 11 Gyr with the assumed \Z,
helium content 0.40 and a basic mass loss of 0.165 \Msun\ (for the case
Y=0.25), so we assume a milder dependence of the mass loss on the
evolving red giant mass. Another constraint is given by the necessity of
keeping the required number of stars within the red clump (and with the
observed distribution in luminosity), since, with increasing mass loss,
the evolutionary tracks tend to shift to the blue (see Fig. \ref{fig3}).

\begin{figure}
   \centering
   \includegraphics[width=8cm]{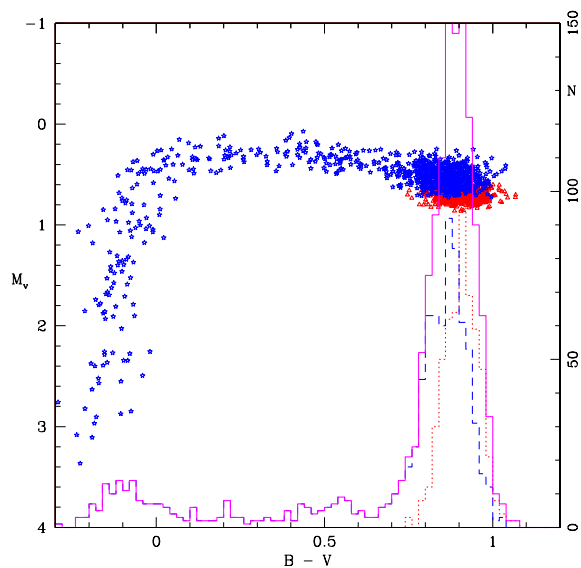}
      \caption{A synthetic CM diagram of the HB in NGC 6441,
      for an age of 11Gyr, mass loss of 0.165\msun with $\sigma=0.02$ and
      extra mass loss as in the text.
      Stars with
      fixed Y=0.25 (open triangles, red) plus stars with variable Y up
      to 0.40 (stars, blue). For the helium distribution, see text and
      the following figures. The standard deviation assumed for both V
      and \BMV\ in this and in the following figures is 0.05mag.
	  The histograms give the number of stars as a function of B--V: the total
	  number is represented by the full line histogram, while the dotted on 
	  is the Y=0.25 population and the dashed one is the population with
	  variable Y. 
	  }
         \label{fig5}
   \end{figure}

\begin{figure*}
   \centering
   \includegraphics[width=14cm]{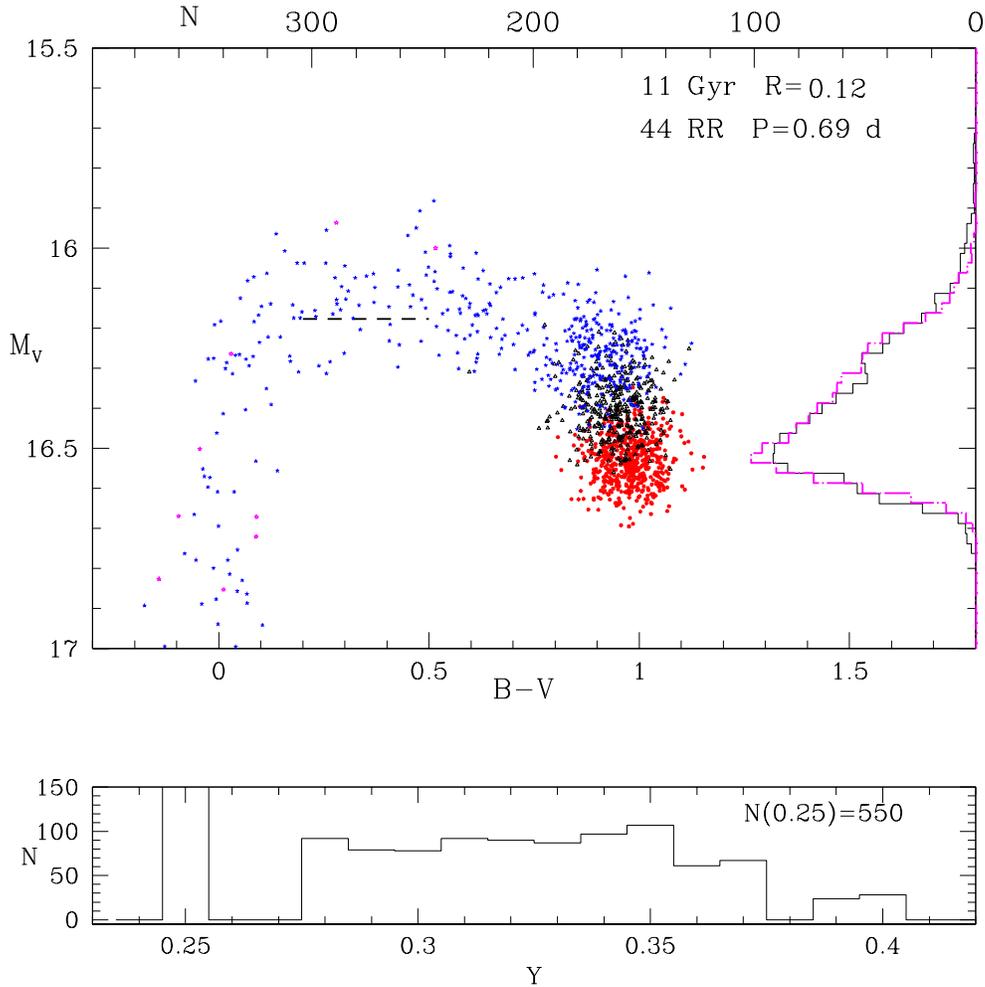}
      \caption{
      For the simulation of Fig. 5, we show in the top panel the
      synthetic CM of the red clump region, with the details of the
      distribution in luminosity (line-dotted histogram, magenta). The
      continuous line gives the observed histogram, as in
      Fig. \ref{fig2}. Different symbols (and
      colours) denote different helium contents. Points: stars with
      Y=0.25 (red); open triangles (black): 0.27$<$Y$<$0.32; 
      asterisks (blue): 0.32$<$Y$<$0.38;
      stars (magenta): 0.38$<$Y$<$0.40. The dashed line indicates the
      approximate position of the RR Lyr variables, derived from 
      Pritzl et al. 2003. The assumed age, the number
	  of RR Lyr (RR) and the average period of the RR Lyr (P), 
	  and the fraction of HB stars bluer than \BMV\ =
      0.5 (R) in the simulation  are listed at the top right. 
	  The lower panel shows the number distribution
      of HB stars vs. helium content corresponding to the CM diagram in
      the upper panel. The total number of the first stellar generation
	  with Y=0.25 is given at the top right and labeled N(0.25).}
         \label{fig6}
   \end{figure*}

In Fig. \ref{fig5} and Fig. \ref{fig6} we give an example of a synthetic
HR diagram and of its detailed behaviour in the clump region.  In
addition to the parameters specified above---age 11 Gyr, $\Delta$M=0.165
\Msun, $\sigma$=0.02---, the extra mass loss included takes the form
$0.10 \times[M_{rg}(Y=0.25)-M_{rg}(Y)]$. Here we discuss the main
features of these diagrams.

To reproduce the strong upward tilt in the red clump, it is necessary to
consider objects with helium content up to Y $\sim$ 0.35. In
Fig. \ref{fig10} the typical relation of \BMV\, with Y is shown. To get
\BMV\ $<$ 0.5 mag, one needs stars with Y $>$ 0.33, while the main body
of clump stars (\BMV\, $>$ 0.5) have a Y content up to 0.37; to find
(almost) exclusively blue stars we have to attain Y $>$ 0.37.

In order to have, on the one side, blue HB stars at least $\sim$ 3 mag
less luminous than the HB luminosity peak, and on the other, the
expected number of red clump members, it is necessary to assume a non
uniform Y distribution for the second star
generation. Otherwise, it is not possible to obtain a blue tail
without remaining with too few red stars (\BMV\ $>$ 0.5).

\begin{figure}
   \centering
   \includegraphics[width=8cm]{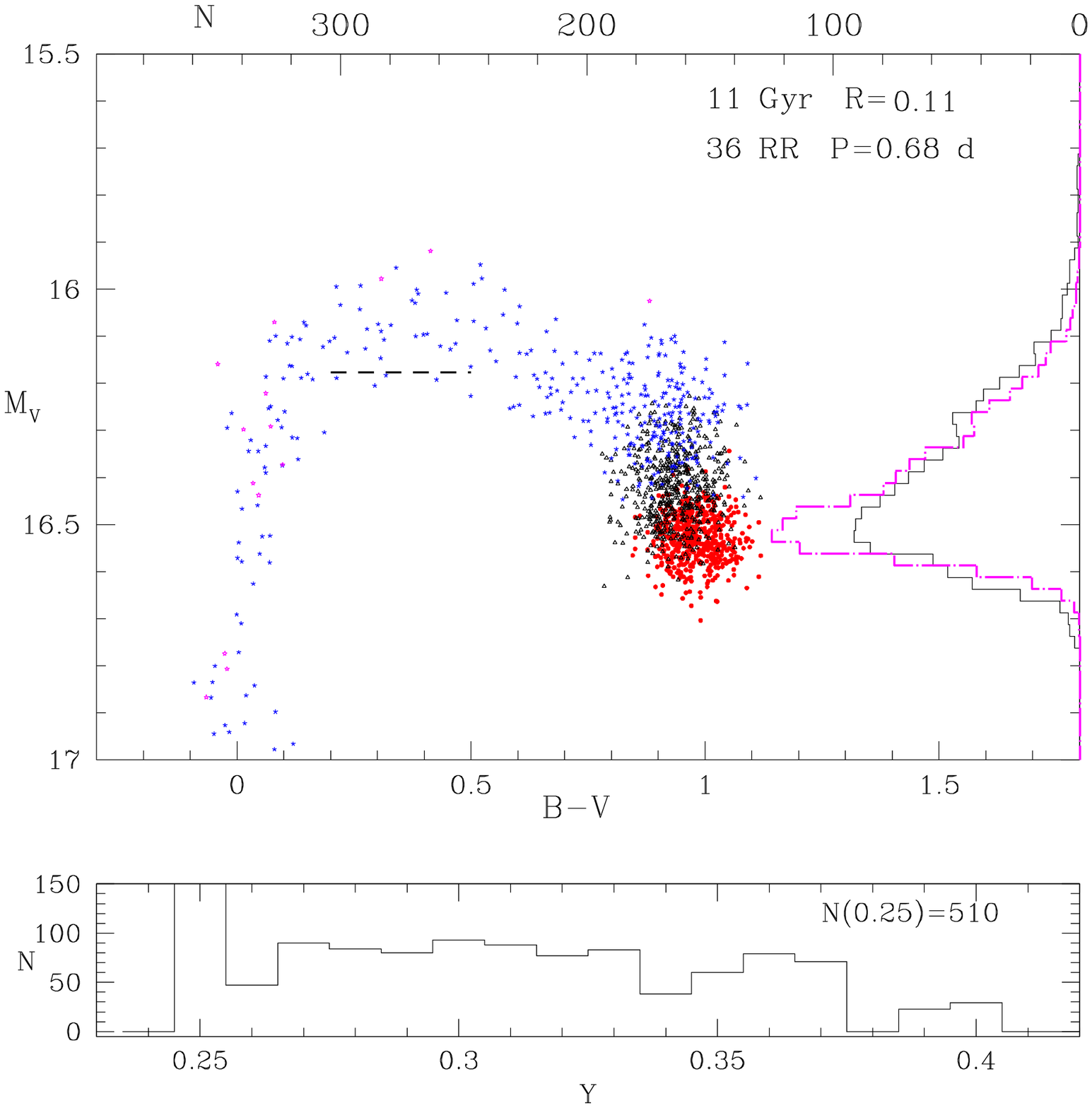}
      \caption{As in Fig. \ref{fig6}, but with no gap in the Y
      distribution (see the lower panel): it is not possible to avoid a
      peak in the luminosity histogram substantially larger than the 
      one observed}
         \label{fig7}
   \end{figure}

\begin{figure}
   \centering
   \includegraphics[width=8cm]{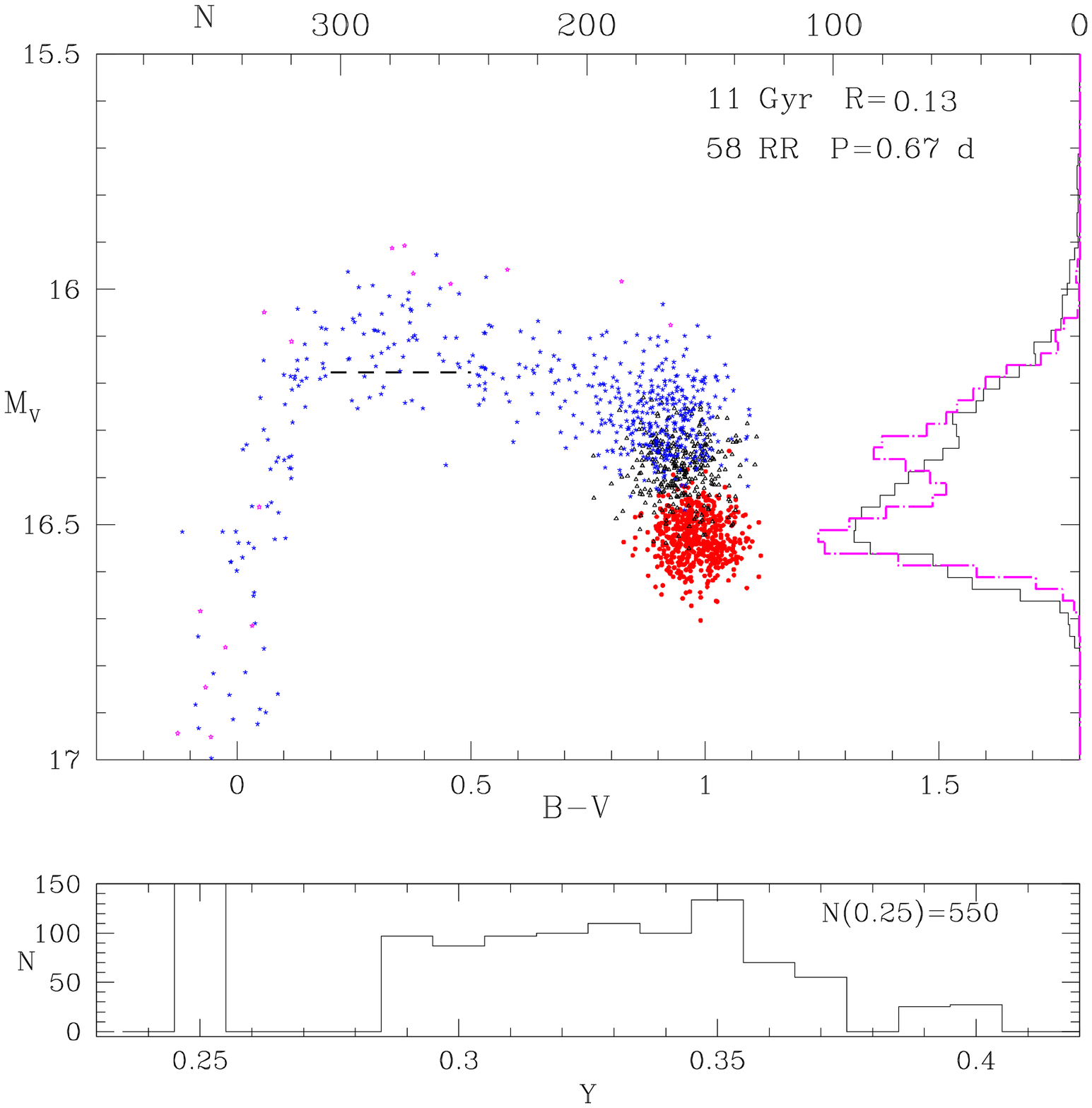}
      \caption{As in Fig. \ref{fig6}, but with a gap of 0.035 between
    the Y of the red clump and the lower limit of the population with 
    varying Y: two well defined peaks appear in the luminosity histogram.}
         \label{fig8}
   \end{figure}

The difference in helium content between the first and the second
generations has to be $\sim$ 0.02 (a gap in Y from 0.25 to 0.27), in
order not to increase the peak in the clump distribution 
(Fig. \ref{fig7}). At the same time, the gap cannot be
larger than 0.03 (Y from 0.25 to 0.28), otherwise the clump distribution
would show (always) two well defined peaks (Fig. \ref{fig8}).

\begin{figure}
   \centering
   \includegraphics[width=8cm]{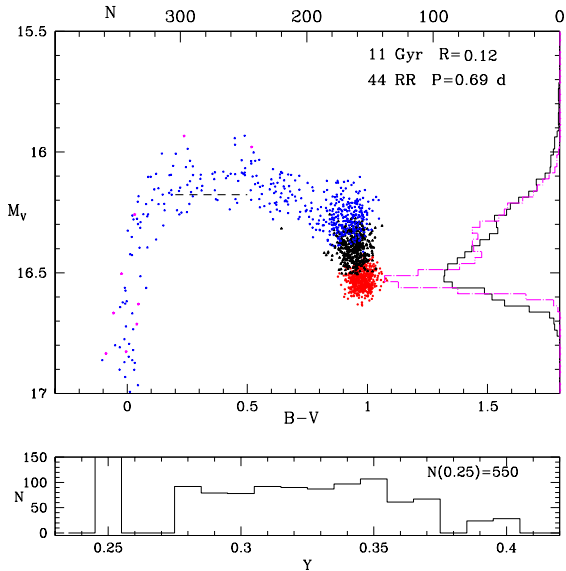}
      \caption{As in Fig. \ref{fig6}, except that the standard
      deviation on V and \BMV\ is
      reduced from 0.05 to 0.03 mag: the clump appears separated in two
      groups.}
         \label{fig9}
   \end{figure}

The importance of the assumed error on \V\,,\BMV\, is exemplified in
Fig. \ref{fig9}, where the same simulation of Fig. \ref{fig6} is shown with
an error on \V\,,\BMV\, reduced from 0.05 to 0.03 mag: the clump appears
clearly separated in two groups. This illustrates the relevance of
improving the photometric precision in order to understand better the
population stratification in GCs.

\subsection{Metallicity}

The adopted metallicity Z=0.006 falls on the lower side of the values
acceptable for the cluster under exam \citep[e.g.][]{clementini2005}. In
fact, it corresponds to the value in the \cite{grattoncarretta} scale,
if we do not consider $\alpha$--enhancement. On the other hand, this
metal rich cluster seems to be also $\alpha$--enhanced (Carretta,
private communication).  Should Z be larger, all the estimated helium
values would be pushed towards even larger values.


\subsection{``The best solution''}

We consider a satisfactory result the one given by the simulation in
Fig. \ref{fig5} and Fig. \ref{fig6}. The helium distribution in shown in
the lower panel in Fig. \ref{fig6}; an extra mass loss of $0.10 \times
[M_{rg}(Y=0.25)-M_{rg}(Y)]$ has been assumed, beside the standard
$\Delta$M=0.165 \Msun\, mentioned before. The simulation has the
correct value of the fraction of stars at \BMV\ $<$ 0.5 mag
within the errors, a number of RR Lyr
variables close to the estimated one (45 in the sample in Piotto et al.,
as mentioned before) with a mean period of 0.69 ~d close to the observed
fundamentalized period of 0.67~d.  As for the very blue stars, three
stars are found at \Mv\ $>$ 3.0 mag in Fig. \ref{fig5}.

We stress again that variables and BHB stars are only a tail of
the cluster star distribution, and that a small fluctuation in the clump
number ($\pm \sqrt(N_{clump})$ and in the number of stars which leave
the clump at a luminosity slightly lower than the average, changes
noticeably the number of RR Lyraes and their mean period.

\subsection{Dispersion in mass loss and maximum helium content}

In the simulations we adopt $\sigma=0.02$\, a widely used value in
the relevant literature (\cite{lee1994,catelan2001}). With this assumption,
it is necessary to include stars with Y up to 0.40. It is important to see 
whether such extreme values of Y can be avoided, since current models for
possible self--enrichment in helium in globular clusters do not 
predict valus larger than Y$\simlt 0.36$. We performed simulations with 
larger values for $\sigma$: while it is possible to achieve the end of the
blue tail with Y$<$0.4, the mean RR Lyr period requires the
presence of stars with Y up to $\sim 0.37 - 0.38$. These stars
in fact have the (large) luminosity  necessary to obtain the long 
observed periods. 


\subsection{The number vs. helium distribution}

\begin{figure}
   \centering
   \includegraphics[width=8cm]{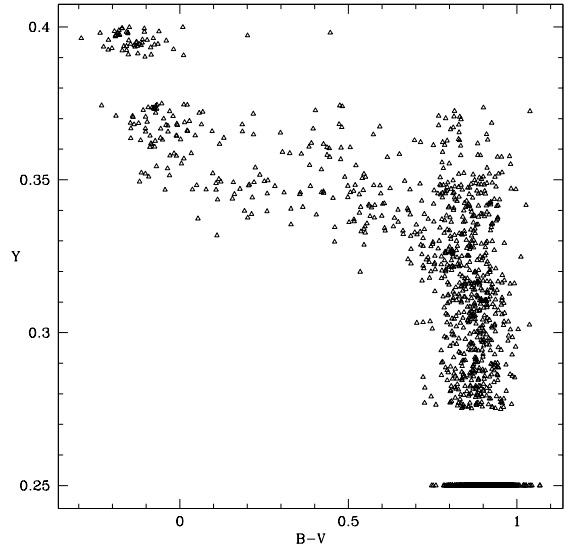}
      \caption{The relation between \BMV\ and helium content Y for the
    case in Fig. \ref{fig6}, but a quite similar behaviour has been
    found in all the simulations.}
         \label{fig10}
   \end{figure}

The number of stars as function of Y assumed in the simulations is given 
in the bottom panels of Fig. 6, 7 and 8. For our ``best solution" we see
that only 550/1452$\sim$38\% of stars belong to the first stellar generation
with ``normal" helium Y=0.25. Anoher
important feature is made evident by Fig. \ref{fig10}. Stars with Y
$>$ 0.33 are necessary to achieve \BMV\ $<$ 0.5 mag, but a large
fraction of these objects remains in the red clump, due to the large
excursions in colour of helium rich structures (see
Fig. \ref{fig3}). Besides, the presence of structures with Y up to $\sim$
0.37 is required by the luminosity distribution of the clump itself
(Fig. \ref{fig6}). This means that we must have a substantial amount of
stars with Y $>$ 0.33 in the cluster. In fact, in our ``best solution''
(\S 4.3 and Fig. 5), we have $\sim$38\% of stars with Y = 0.25, $\sim$33\% of
stars with 0.275$\leq Y \leq$0.33, and $\sim$29\% of stars with Y $>$
0.33. About 15\% of stars have Y $>$ 0.35. In this last group 
are included the objects with Y = 0.39 -- 0.40
(3.6\%), which are the only ones to remain (almost) always at \BMV\
$\leq$ 0.0.

The exact numbers may be debated, as they depend in part on the age and
mass loss along the RGB, and the extreme range Y = 0.39 -- 0.40 can be
avoided with a mass dispersion $\geq$ 0.04 \msun, but surely a
substantial fraction ($\approx$ 30\%) has Y $>$ 0.33.

\section{The R parameter and the bump location}

The ratio R of the number of HB stars to the number of red giants above
the HB level has been a powerful indicator of the helium content in
cluster stars, as first suggested by \cite{ibenNature}, when the aim was 
to discriminate the helium mass
fraction in the oldest stars of the Galaxy between values close to Y=0.1
or to Y=0.3. For NGC 6441, Salaris et al. (2004) find a value R=$1.85 \pm
0.11$, consistent with the theoretical expectation 
by the same authors for the
cluster metallicity.  On the contrary, the simulations by
\cite{sweigart1998} predicted values as large as 3.4 -- 3.9, when they
assumed main sequence helium abundances
of Y=0.38 and Y=0.43 for all the cluster stars. For these values,
the level of luminosity of the HB clump is
highly enhanced and thus the number of red giants to be considered
reduced. This in fact is also predicted by the simple interpolation
formula by Buzzoni et al. (1983)
\begin{equation}
Y=0.380 \log R +0.176
\end{equation}
We do not explicitly compute the value of R from our simulations, because
it is easy to understand that we would find a ``normal'' value for R, due to
the fact that the HB level in the R definition is the zero age value, which
in this cluster is determined by the stars having the normally
adopted value of Y=0.25. These are the lowest 
luminosity stars in the red clump and so the number of red giants to
be considered remains the same as in the case with constant Y=0.25. 
In addition, there is only a small
percentage of very hot stars with very high helium, whose HB lifetimes would
exceed the Y=0.25 lifetimes, so that the correction to R due to these
stars would turn out to be small.
Therefore, the argument by \cite{layden1999} based on the R parameter
can not be considered a proof against a high helium content 
as explanation of the RR Lyr long periods. 

\cite{raimondo2002} notice that a uniform high helium abundance of the
cluster stars predicts a wrong location of the red giant branch bump 
\citep{thomas1967, iben1968} with respect to the HB. Also in this
case, our modeling is consistent with this latter parameter:
a prominent RGB bump will occur at the luminosity predicted by 
models with Y=0.25, which is the abundance of $\sim$40\% of the stars. 
The stars having larger Y will produce smaller bumps, spread along the RGB
at increasing luminosities, according to the number vs. Y distribution.
Notice also that the prominence of the bump decreases with increasing Y.
Further, the HB reference luminosity level in our simulations, as 
mentioned above, corresponds to the Y=0.25 level. Therefore,
the quoted difficulty with a high helium population by  \cite{raimondo2002} 
does not apply to our model.

\section{Discussion}

The interpretation of the HB morphology of NGC~6441 in terms of a helium
rich population has provided a coherent interpretation of the main
features of the anomalous HB of this cluster, namely, the thick
extension in luminosity of the red clump, the very long mean period of
the RR Lyrs for a cluster of high metallicity, the extension into the
blue of the HB. 
Modellization shows that a continuous distribution of helium
content above Y=0.25 is not consistent with the luminosity distribution
of stars in the red clump, and that a helium discontinuity {\it must} be
present, from Y=0.25 to a minimum value of Y$\simeq$0.27, like in the
cluster NGC~2808. {\it We suggest that a more precise photometry of the
HB would put into evidence a double peak in the luminosity distribution
of the red clump stars}.

While it had been long suspected that the blue HB and the high RR Lyr
luminosity in this cluster ---and in its twin NGC 6388--- could be
easily attributed to a high helium content in the small fraction of
stars --$\sim 12$\%-- constituting the blue HB \citep{sweigart1998,
dantonacaloi2004}, it comes as a surprise of these detailed simulations
that a much more important fraction of stars has a very large Y: 29\%
have Y $>$ 0.33 and $\sim$14\% have Y $>$ 0.35. This latter figure is
similar to the percentage of stars with Y $\sim$ 0.4 found from the
analysis of the main sequence in NGC 2808 \citep{dantona2005}. In this
respect, we conclude that both NGC 6441 and NGC 2808 were able to form
about the same fraction of very high helium stars, whatever their
origin.

The difference between the two clusters resides in the {\it intermediate
helium} population, which for NGC 2808 contains 35\% of stars, and
covers only the range between 0.26 and 0.29 \citep{dantona2005}. In NGC
6441, this fraction is $\sim$ 48\% and covers a helium content from 0.27
to 0.35.  Thus, any dynamical and chemical evolution model for these
clusters must also explain the difference in this intermediate
population. We suggest that it is due to the different metallicities of
the two clusters: in NGC 2808 (Z $\sim 1-2 10^{-3}$) the maximum mass
for AGB type evolution is smaller than in NGC 6441 (Z
$\simgt 6 10^{-3}$). Therefore, the range of progenitor masses for the
second stellar generation in NGC 6441 contains stars of larger mass
which also have larger helium content \citep{ventura2002}.

One of the motivations to pursue the present work was to investigate
whether an ``extreme'' high helium population is present in other
clusters beside NGC 2808 and $\omega$Cen. We found this extreme helium
population in NGC 6441 (and this must also exist in the very similar
cluster NGC 6388), and we also discovered that more than 60\% of the
cluster stars belong to the second stellar generation, having
$0.27\simlt Y \simlt 0.4$.

The interpretation of the HB morphology of NGC 6441 makes it even more
cogent to understand better which is the origin of these extreme helium
rich stars.

\begin{acknowledgements}
We thank dr. Anna Marenzi for help in the preparation of the paper and
prof. G. Piotto for providing the database of NGC 6441.
\end{acknowledgements}


%


\end{document}